\documentclass[aps,pre,superscriptaddress,showpacs,twocolumn,floatfix]{revtex4-1}
\usepackage{epsfig}
\usepackage{color}

\usepackage{tikz}
\usepackage[normalem]{ulem}
\usepackage{amsmath}
\usepackage{amssymb}
\usepackage{placeins}

\definecolor{red}{rgb}{1,0,0}
\definecolor{blue}{rgb}{0,0.1,0.9}
\definecolor{black}{rgb}{0,0,0}

\def\fv{\vec{f}}
\def\Fv{\vec{F}}
\def\rv{\vec{r}}
\def\Uv{\vec{U}}
\def\Ic{{\cal I}}

\def\Lc{{\cal L}}
\def\Dc{{\cal D}}

\def\edot{\dot\epsilon}
\newcommand{\p}{\partial}
\newcommand{\eq}[1]{\begin{align}#1\end{align}}

\newlength{\arrow}
\def\rv{\vec{r}}

\def\nv{\vec{n}}

\def\Vv{\vec{V}}

\def\uv{\vec{u}}
\def\omv{\vec{\omega}}
\def\tauv{\vec{\tau}}

\def\ij{\langle i j \rangle}

\begin{document}

\title{Phase Diagram for Inertial Granular Flows}
\author{E. DeGiuli}
\affiliation{New York University, Center for Soft Matter Research, 4 Washington Place, New York, NY, 10003}
\affiliation{Institute of Theoretical Physics, Ecole Polytechnique F\'ed\'erale de Lausanne (EPFL), CH-1015 Lausanne, Switzerland}
\author{J.N. McElwaine}
\affiliation{Department of Earth Sciences, Durham University, Science Labs, Durham, DH1 3LE, U.K.}
\author{M. Wyart}
\affiliation{Institute of Theoretical Physics, Ecole Polytechnique F\'ed\'erale de Lausanne (EPFL), CH-1015 Lausanne, Switzerland}


%

\begin{abstract}


 Flows of hard granular materials depend strongly on the interparticle friction coefficient $\mu_p$ and on the inertial number $\Ic$, which characterizes proximity to the jamming transition where flow stops. Guided by numerical simulations, we derive the phase diagram of dense inertial flow of spherical particles, finding three regimes for $10^{-4} \lesssim \Ic \lesssim 10^{-1}$: \textit{ frictionless, frictional sliding, } and {\it rolling}. These are distinguished by the dominant means of energy dissipation, changing from collisional to sliding friction, and back to collisional, as $\mu_p$ increases from zero at constant $\Ic$. The three regimes differ in their kinetics and rheology; in particular, the velocity fluctuations and the stress ratio both display non-monotonic behavior with $\mu_p$, corresponding to transitions between the three regimes of flow. We rationalize  { the phase boundaries between these regimes}, show that energy balance yields scaling relations { between microscopic properties} in each of them, and { derive the strain scale at which particles lose memory of their velocity. For the frictional sliding regime most relevant experimentally, we find for $\Ic\geq 10^{-2.5}$ that the growth of the macroscopic friction $\mu(\Ic)$ with $\Ic$ is induced by an increase of collisional dissipation. This implies in that range that $\mu(\Ic)-\mu(0)\sim \Ic^{1-2b}$, where $b\approx 0.2$ is an exponent that characterizes both the dimensionless velocity fluctuations ${\cal L}\sim \Ic^{-b}$ and the density of sliding contacts $\chi\sim \Ic^b$. }



\end{abstract}
\maketitle

Dense flows of granular media are central to many industrial processes and geophysical phenomena, including landslides and earthquakes \cite{Gennes99,Andreotti13,Nedderman92}. At a fundamental level,  describing such driven materials remains a challenge, in particular near the jamming transition where crowding effects become dominant and flow stops. 
In the last decade, progress was made by considering  the limit of {\it perfectly rigid} grains,
for which dimensional analysis implies that the strain rate $\edot$, the pressure $P$ and the grain density $\rho$ can only affect flows via the inertial number $\Ic = \edot D\sqrt{\rho/P}$, where $D$ is grain diameter \cite{MiDi04,Cruz05,Lois05}. In particular, for stationary flows the packing fraction $\phi$ and stress anisotropy $\mu= \sigma/P$, where $\sigma$ is the shear stress, are functions of $\Ic$. From the constitutive relations $\phi(\Ic)$ and $\mu(\Ic)$ the flow profile can be explained in simple geometries \cite{MiDi04,Forterre08,Sun11,Azema14}. Here we focus on dense flows $\Ic \lesssim 0.1$ for which the networks of contacts between grains span the system and particle motion is strongly correlated \cite{Radjai02,Pouliquen04}, and do not consider the quasi-static regime $\Ic \lesssim 10^{-4}$ where flow appears intermittent \cite{Cruz05,Kruyt07,Gaume11,Henkes16}. 
%
In this intermediate range one finds
\eq{
\label{0}
\mu(\Ic) & = \mu_c + a_\mu \;\Ic^{\alpha_\mu}, \quad \phi(\Ic) = \phi_c - a_\phi \;\Ic^{\alpha_\phi},
}
where $\mu_c$ and $ \phi_c$ are non-universal and depend on details of the grains. Experiments on glass beads and sand find exponents $\alpha_\mu \approx \alpha_\phi \approx 1$, consistent with   numerical simulations using frictional particles reporting $\alpha_\mu = 0.81$ and $\alpha_\phi = 0.87$ \cite{Peyneau09}. Despite their importance, constitutive laws Eq.\ref{0} remain empirical. 
Building a microscopic framework to explain them would shed light on a range of debated issues, including transient phenomena \cite{Andreotti13,Bi11}, non-local effects \cite{Bouzid13,Henann13,Kamrin14}, and the presence of S-shaped flow curves when particles are soft \cite{Otsuki11,Ciamarra11,Grob14}.

\begin{figure}[t!] 
\includegraphics[width=0.48\textwidth,viewport=15 5 230 100,clip]{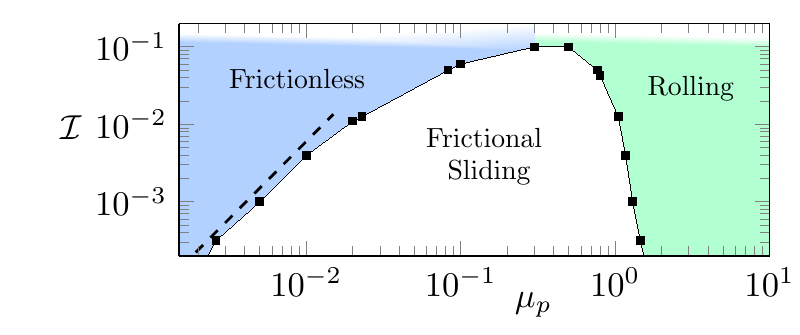}
\caption{(Color online) Phase diagram of dense homogeneous inertial frictional flow. In the frictionless and rolling regimes, most energy is dissipated by inelastic collisions, while in the frictional sliding regime energy dissipation is dominated by sliding. Along the phase boundary, grains dissipate equal amounts of energy in collisions and in sliding. For $\Ic \gtrsim 0.1$, one enters the dilute regime \cite{Azema14}. The dashed line has slope $2$.\label{phase}}
\end{figure}

To make progress, it is natural to consider the limiting case where particles are {\it frictionless}, a situation that has received considerable attention in the jamming literature \cite{Ohern03,Wyart05b,Liu10,Hecke10}. For hard particles, two geometrical results key for inertial flows are as follows. First, as the density increases, the network of contacts becomes more coordinated, implying that motion becomes more constrained. This leads to a divergence of the velocity fluctuations $\langle \delta V \rangle$ when constraints are sufficient to jam the material \cite{Lerner12a,During13,During14,Andreotti12}. Thus the contact network acts as a lever, whose amplitude is characterized by the dimensionless number $\Lc \equiv \langle \delta V \rangle/(\edot D)$. At the same time, the rate at which new contacts are made increases, and the creation of each contact affects motion on a growing length scale. These effects imply that velocity fluctuations decorrelate on a strain scale $\epsilon_v$ that vanishes at jamming \cite{DeGiuli15a}.
The theory of Ref. \cite{DeGiuli15a}, which uses the fact that dissipation can only occur in collisions for frictionless particles, predicts $\alpha_\mu=\alpha_\phi=0.35$, $\Lc\sim \Ic^{-1/2}$ and $\epsilon_v\sim \Ic$. Encouragingly, these results agree with the numerics of Ref. \cite{Peyneau08}, which found $\alpha_\mu \approx \alpha_\phi \approx 0.38$ and $\Lc\sim \Ic^{-0.48}$. However, $\alpha_\mu$  and $\alpha_\phi$ differ significantly from their values for frictional grains stated above, suggesting the presence of different universality classes. Currently, why friction qualitatively affects flows and potentially leads to several universality classes, how many universality classes exist, and what differs between them microscopically are unresolved questions.

\begin{figure}[t!] 
\includegraphics[width=0.46\textwidth,viewport=20 0 650 355,clip]{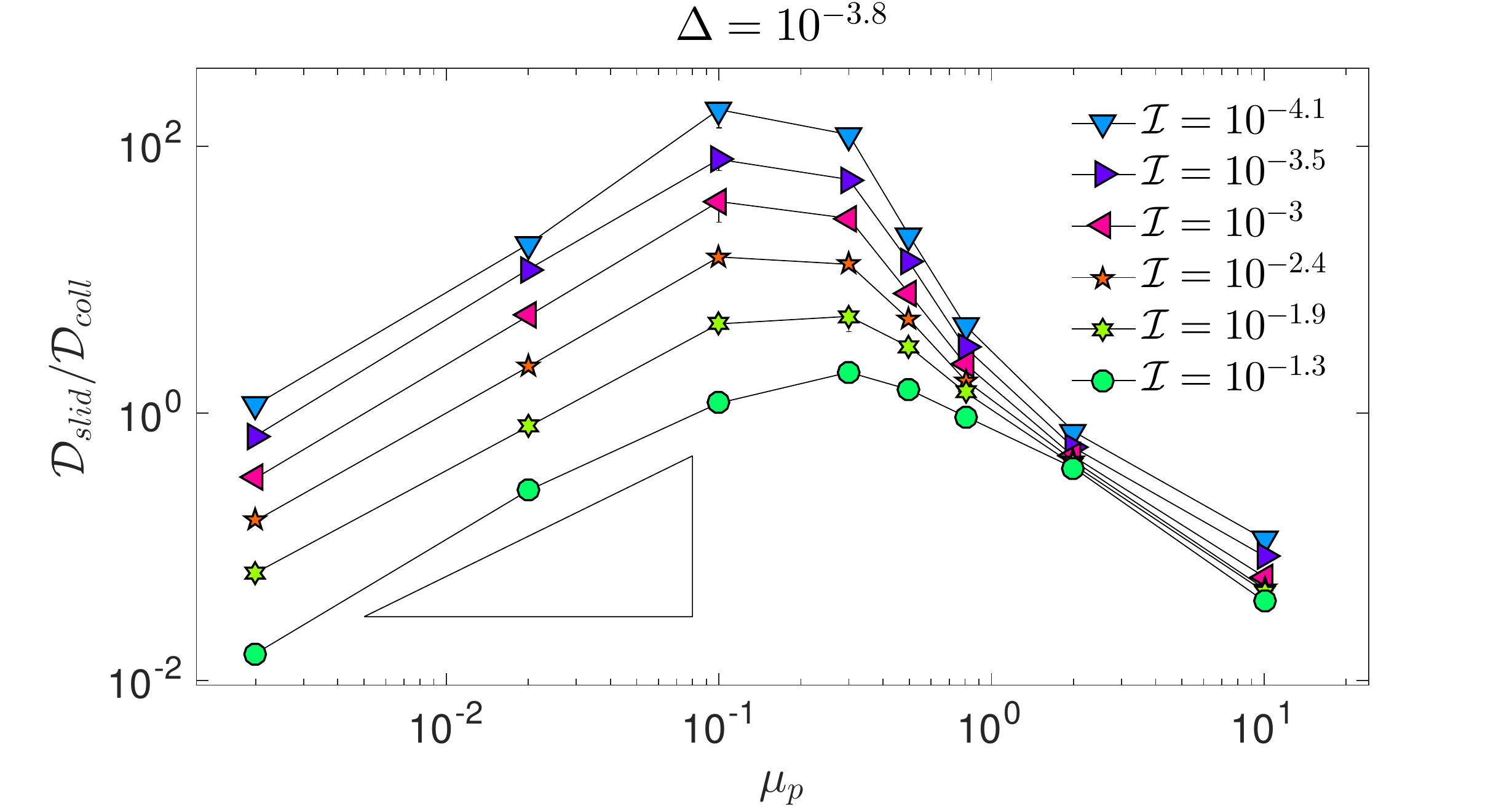}
\caption{(Color online) Ratio of dissipation due to sliding, $\Dc_{slid}$, to dissipation from collisions, $\Dc_{coll}$, vs $\mu_p$. The triangle has slope $1$. 
}\label{dratio}
\end{figure}
In this work we use numerical simulations to answer these questions. We systematically study dense flows over a large range of $\Ic$ and $\mu_p$. By focusing on the microscopic cause of dissipation, we show the existence of three universality classes, as illustrated in Fig.~\ref{phase}.~At low friction, there exists a frictionless regime in quantitative agreement with the theory of Ref. \cite{DeGiuli15a}, in particular we establish that $\epsilon_v\sim \Ic$.  As the friction increases, one enters the {\it frictional sliding} regime, where dissipation is dominated by sliding at contacts instead of collisions, and for which $\epsilon_v\sim \Ic$ holds true but $\Lc\sim \Ic^{-b}$ with $b\approx 0.22$. { We relate the exponent $b$ to the density of sliding contacts, $\chi\sim {\cal I}^b$. Most importantly, we show that although the value of $\mu_c$ in Eq.\ref{0} reflects sliding dissipation, the dependence of $\mu$ with $\Ic$ is governed by collisional dissipation when $\Ic\geq 10^{-2.5}$, leading to $\alpha_\mu=1-2b$.}
 Finally, at even larger $\mu_p$ one enters a  {\it rolling} regime where dissipation is once again dominated by collisions, and where exponents are consistent with those of frictionless particles, both for kinetic observables and constitutive laws. 
We derive the phase boundary between the frictionless and the frictional sliding regime. Overall, our work explains why friction qualitatively changes physical properties, and paves the way for a future comprehensive microscopic theory of dense granular flows.

\textit{Numerical Protocol---} To model inertial flow of frictional particles, we use a standard discrete element method \cite{Cundall79} in two dimensions, described in more detail in Appendix 1. Collisions are computed by modeling grains as stiff viscoelastic disks: when grains overlap at a contact $\alpha$, they experience elastic and viscous forces $\fv_\alpha^{e}$, and $\fv_\alpha^{v}$, respectively, leading to a restitution coefficient which we choose to be $e=0.1$ \footnote{The choice of restitution coefficient has little effect on flow in the dense regime, see \cite{MiDi04,Cruz05,Peyneau08,Peyneau09,Chialvo12,Hurley15}.}. The tangential (normal) components $\fv_\alpha{}^{\!T} \; (f_\alpha^N)$ are restricted by Coulomb friction to satisfy $|\fv_\alpha{}^{\!T}| \leq \mu_p f_\alpha^N$; contacts that saturate this constraint are said to be {\it sliding}, { while those that obey a strict inequality are said to be {\it rolling}. }

Shear is imposed with rough walls bounding the upper and lower edges on an $x-$periodic domain. We perform our numerics at imposed global shear rate and constant pressure, following a system preparation described in Appendix 1. { We discard data that do not satisfy strict criteria for homogeneity of the flow, as specified in Appendix 1. } Grain stiffness is such that relative deformation at contacts is $\Delta\approx 10^{-3.8}$, within the rigid limit established previously \cite{Cruz05}, and system size is large enough to ensure the absence of finite-size effects. Independence of our results with respect to $\Delta$, $e$, and $N$ is shown in Appendix 2.
\begin{figure}[t!] 
\includegraphics[width=0.5\textwidth,clip]{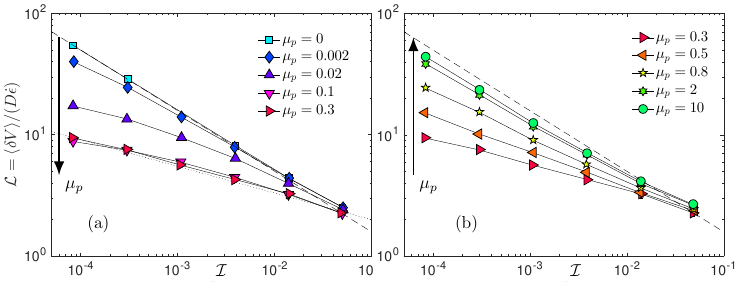}
\caption{(Color online) Lever $\Lc$ vs $\Ic$, (a) for $\mu_p \leq 0.3$, and (b) for $\mu_p \geq 0.3$. The dashed lines are $\propto \;\Ic^{-1/2}$, while the dotted line is $\propto \Ic^{-0.22}$. 
}\label{lever1}
\end{figure}
\textit{Partitioning dissipated power---} Frictional particles can dissipate energy either through inelastic collisions, at a rate $\Dc_{coll}$, or by sliding at frictional contacts, at a rate $\Dc_{slid}$. In our contact model, inelasticity is due to the viscous component of contact forces; therefore the collisional dissipation rate $\Dc_{coll}$ can be written
\eq{ \label{dcoll1}
\Dc_{coll} \equiv \sum_{\alpha \in C} f_\alpha^{N,v} U^N_\alpha + \sum_{\alpha \in C_{\mathrm{R}} } \fv_\alpha{}^{\! T,v} \cdot \Uv_\alpha^{T},
}
where $\Uv_\alpha$ is the relative velocity at contact $\alpha$, decomposed into normal and tangential components, $U^N_\alpha$ and $\Uv_\alpha^{T}$. Here $C$ denotes all contacts, of number $N_c$, and $C_{\mathrm{R}}$ denotes rolling contacts. The dissipation rate due to sliding is
\eq{ \label{dslid1}
\Dc_{slid} \equiv \sum_{\alpha \in C_{\mathrm{S}}} \fv_\alpha{}^{\! T} \cdot \Uv_\alpha^{T},
}
where $C_{\mathrm{S}}$ is the set of sliding contacts. In steady state, dissipation must balance the work done at the boundaries \cite{Chaikin00}. The energy input from the shear stress is $\Omega \sigma \edot$, where $\Omega$ is the system volume, and for large systems, additional contributions from fluctuations of the normal position of the wall are insignificant. { We  define dimensionless dissipation rates per particle $\tilde\Dc_{coll}\equiv\Dc_{coll}/(\Omega p \edot)$, $\tilde\Dc_{slid}\equiv\Dc_{slid}/(\Omega p \edot)$, so that \cite{Hurley15} 
\eq{ \label{energy}
\mu = \tilde\Dc_{coll} + \tilde\Dc_{slid}.
}
}
To investigate which source of dissipation dominates in Eq.~\ref{energy}, we consider the ratio $\Dc_{slid}/\Dc_{coll}$, shown in Fig.\ref{dratio}. As expected, collisional dissipation dominates in the frictionless limit, but sliding dissipation becomes more important as $\mu_p$ is increased, and becomes {\it dominant} at intermediate friction coefficients and small inertial number, consistent with earlier simulations for $\mu_p=0.3$~\cite{Hurley15}. Strikingly,  the dependence on $\mu_p$ is non-monotonic: when $\mu_p$ reaches $\approx 0.2$, this trend abruptly reverses, and $\Dc_{slid}/\Dc_{coll}$ {\it decreases} with $\mu_p$, implying that collisional dissipation dominates as $\mu_p \to \infty$. 


To define phase boundaries, we use the inertial number at which $\Dc_{slid}/\Dc_{coll}=1$, resulting in the phase diagram of Fig.~\ref{phase}. From the non-monotonicity of $\Dc_{slid}/\Dc_{coll}$ with $\mu_p$, this leads to two phase boundaries merging at $\Ic \approx 0.1$, where the dense flow regime ends \cite{Azema14,Hurley15}. This defines {\it three} flow regimes: {\it frictionless, frictional sliding, } and {\it rolling}, where sliding dissipation dominates only in the intermediary regime. Later in this work, we will show that this phase diagram correctly classifies kinetics as well as constitutive laws.

\textit{Connecting dissipation to key kinetic observables---} 
%
%
In the rigid limit, collisions become very short in duration, and the power dissipated in collisions can be expressed in terms of microscopic observables \cite{DeGiuli15a}, as we now recall. 
 Each time a particle changes its  direction with respect to its neighbors, a finite fraction of its kinetic energy $\sim m \delta V^2$ must be dissipated, where $m$ is the particle mass (we consider finite restitution $e<1$). Since $\epsilon_v$ is the characteristic strain at which velocities decorrelate, this occurs at a rate $\propto \edot/\epsilon_v$, thus $\Dc_{coll}\sim N (\edot/\epsilon_v) m \langle \delta V^2\rangle$ and
\eq{ \label{dcoll2}
\tilde\Dc_{coll} \propto \frac{N (\edot/\epsilon_v) m \delta V^2}{N D^d p \edot} \propto \frac{\Ic^2 \Lc^2}{\epsilon_v}
}
\begin{figure}[t!] 
\includegraphics[width=0.48\textwidth,clip]{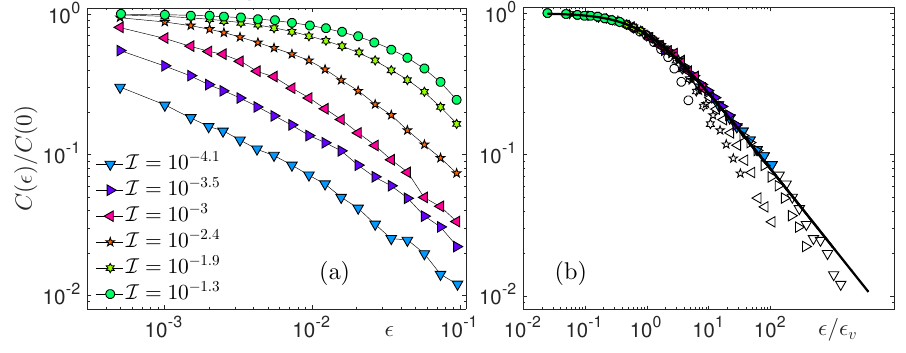}
\caption{(Color online) Autocorrelation of particle velocities, $\tilde C(\epsilon) = \langle V^y_i(0) V^y_i(\epsilon) \rangle/ \langle V^y_i(0)^2\rangle$ for $\mu_p=0.3$ and indicated $\Ic$.
 (b) $\tilde C(\epsilon)$ vs $\epsilon/\epsilon_v$. In (b), unfilled symbols correspond to strains larger than $0.01$, not used for fitting, and the solid line shows the fitted form. }\label{corr}
\end{figure}

The rate of sliding dissipation can be directly estimated from its microscopic expression, Eq.~(\ref{dslid1}). We assume that the force at the sliding contact is typical, \textit{i.e.} $|\fv_\alpha^T| = \mu_p f_\alpha^N \sim \mu_p p D^{d-1}$, and that the sliding velocity is of the order of the velocity fluctuation, i.e. $|\Uv_\alpha^T| \sim \delta V$. These assumptions hold true in the sliding frictional regime where they matter (they eventually break down in the rolling regime where sliding contacts become rare and atypical, see Appendix 3). 
We get the estimate
\eq{ \label{dslid2}
\tilde\Dc_{slid} \propto \frac{N_c \chi \langle |\fv^T| \rangle_{S} \langle |\Uv^T| \rangle_{S}}{N D^d p \edot} \sim \mu_p \chi \Lc,
}
where $\langle \cdot \rangle_{S}$ denotes an average over sliding contacts, whose fraction is $\chi$. 
 Using Eqs.~(\ref{energy},\ref{dcoll2},\ref{dslid2}) we now get the following constraints on the different  regimes:
\eq{
\mu & \sim \Ic^2 \Lc^2 \epsilon_v^{-1} \qquad & \mbox{Frictionless, Rolling} \label{energy1} \\
\mu & \sim \mu_p \chi \Lc, \qquad & \mbox{Frictional Sliding} \label{energy2}
}
We now test these scaling relations and use them to compute the boundary of the frictionless regime. 


\textit{Measuring kinetic observables--} We measure the lever effect defined as  $\Lc \equiv \langle \delta V \rangle/(\edot D)$, 
where $\langle \delta V \rangle$ is the typical magnitude of velocity fluctuation about the mean velocity profile \footnote{In simulations with walls, the mean velocity profile is not linear \cite{Cruz05}, therefore this differs slightly from the non-affine velocity \cite{DeGiuli15a}.}. Our results are shown in Fig.\ref{lever1}. For any $\mu_p$, $\Lc$ grows as $\Ic \to 0$. In the frictionless limit, we find $\Lc \propto \Ic^{-0.50}$, in agreement with earlier results \cite{Peyneau08} and the prediction   \cite{DeGiuli15a}. A striking result is that the amplitude of this growth is non-monotonic in $\mu_p$, with a minimum around $\mu_p \approx 0.2$, thus closely paralleling the phase diagram of Fig.\ref{phase}. Moreover, in the $\mu_p \to \infty$ limit, the divergence is again close to $\Lc \propto \Ic^{-0.50}$. In contrast, curves that are fully in the frictional sliding regime, as occurs for $\mu_p=0.1$ or $\mu_p=0.3$, are well fitted by { $\Lc \propto \Ic^{-b}$ with $b=0.22$}, close to experiments finding $\Lc \propto \Ic^{-1/3}$ \cite{Menon97,Pouliquen04}. 

\begin{figure}[t!] 
\includegraphics[width=0.5\textwidth,clip]{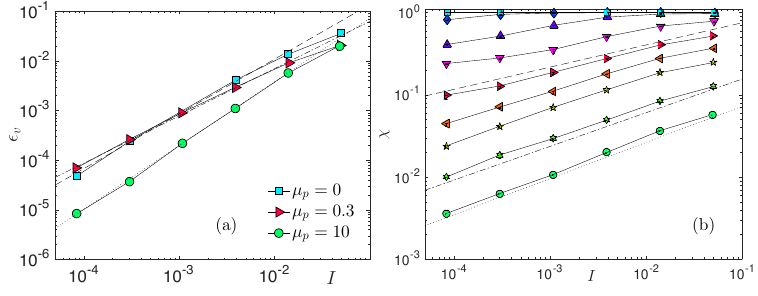}
\caption{(Color online) (a) Decorrelation strain scale, $\epsilon_v$, vs $\Ic$ for selected $\mu_p$. The dotted, dashed, and dot-dashed lines are $\propto \Ic^{1.25}, \Ic^{1.1}$ and $\Ic^{0.9}$, respectively. (b) The fraction of sliding contacts, $\chi$, vs $\Ic$, for various $\mu_p$ (symbols as in Fig. 3). Dotted, dashed, and dot-dashed lines have slopes 0.43, 0.41, and 0.27, respectively. }\label{epsv1}
\end{figure}

We now turn to  the strain scale $\epsilon_v$ beyond which a particle loses memory of its velocity. It can be extracted from the decay of the autocorrelation function \cite{Olsson10a} $C(\epsilon) = \langle V^y_i(0) V^y_i(\epsilon) \rangle$, where we use the vertical component of velocity at particle $i$, $V^y_i$, averaged over all particles and initial time steps. The normalized correlation function $\tilde C(\epsilon) = C(\epsilon)/C(0)$ is shown for $\mu_p=0.02$ and various $\Ic$ in Fig.~\ref{corr}a. We see that beyond a scale $\epsilon_v$, $\tilde C(\epsilon)$ decays as a power-law, as observed numerically in over-damped suspensions \cite{Olsson10a,DeGiuli15a}. For all $\mu_p$, $\tilde C(\epsilon)$ has a similar form; we find that for $\epsilon \lesssim 10^{-2}$ it is well-fitted by $[1+(\epsilon/\epsilon_v(\Ic))^{\nu})]^{-\eta}$, with $\nu=1.1$ and $\eta$ dependent on $\mu_p$. By rescaling $\epsilon$ to obtain a collapse, shown in Fig.~\ref{corr}b, we obtain the scale $\epsilon_v$. Repeating this process for all $\mu_p$ leads to the results shown in Fig.~\ref{epsv1}a for selected $\mu_p$; other $\mu_p$ are shown in Appendix 4. We observe that for all $\Ic$ and all $\mu_p$, we have approximately $\epsilon_v \approx \Ic$, although our best exponent for the rolling regime is closer to $\epsilon_v \approx \Ic^{1.28}$. This result is thus in excellent agreement with the prediction of \cite{DeGiuli15a} for the frictionless case, and also with experimental measurements finding $\epsilon_v \sim \Ic$ \cite{Menon97}. { We recall below our previous argument, which we expect to hold more generally for frictional particles.}



Finally, the fraction of sliding contacts $\chi$ is shown in Fig.~\ref{epsv1}b. For each $\mu_p$, $\chi$ decays as $\Ic \to 0$.  In the frictionless regime, $\chi \approx 1$, as expected, while in the frictional sliding and rolling regimes, $\chi$ decays as a power-law as $\Ic$ is decreased. For the frictional regime, such as  $\mu_p = 0.3$, data are well-fitted by $\chi \sim \Ic^{0.27}$, while in the rolling regime we find a sharper decay, $\chi \sim \Ic^{0.43}$ for $\mu_p=10$. 


Our results on microscopic quantities are summarized in Table~\ref{tab}. We see that the scaling relations Eqs.~(\ref{energy1},\ref{energy2}) are consistent with the data.

\begin{table}[b!]
\begin{tabular}{| c | c || c  | c |}
\hline
\; Regime \; & \; Relation \; & \; Prediction \; & \; Measured \; \\
\hline \hline
Frictionless &  $\Lc \sim \Ic^{-b}$ & $b=1/2$ & $b=0.50$ \\   
 &  $\epsilon_v \sim \Ic^c$ & $c=1$ & $c=1.10$ \\   
\hline
 &  $\Lc \sim \Ic^{-b}$ & & $b=0.22$ \\   
Frictional &  $\epsilon_v \sim \Ic^c$ & { $c=1$ }& $c=0.95$ \\   
Sliding &  $\chi \sim \Ic^{d}$ & $d = b$ & $d=0.27$ \\   
 &  $\delta\mu \sim \Ic^{\alpha_\mu}$ & { $\alpha_\mu = 1-2b$ } & $\alpha_\mu =0.6$  \\   
\hline
 &  $\Lc \sim \Ic^{-b}$ & & $b=0.50$ \\   
Rolling &  $\epsilon_v \sim \Ic^c$ & { $c=1$ } & $c=1.28$ \\   
 &  $\chi \sim \Ic^{d}$ & & $d=0.43$ \\   
\hline
\end{tabular}
\caption{ Summary of scaling behavior. Predictions for the frictionless regime are quoted from the theory of \cite{DeGiuli15a}, while the other predictions are Eqs.~(\ref{energy1},\ref{energy2}). In the frictional sliding regime, scalings are taken for the extremal value $\mu_p=0.3$, while for the rolling regime, scalings are taken from $\mu_p=10$. } \label{tab}
\end{table}

\;

\textit{Regime boundaries --} We can now estimate when the frictionless regime breaks down.  Since in that regime $\Lc \sim \Ic^{-0.5}$, $\epsilon_v \sim \Ic$, and $\chi \sim 1$, we have according to Eqs.~(\ref{dcoll2},\ref{dslid2}) $\Dc_{slid}/\Dc_{coll} \sim \mu_p \Ic^{-0.5}$, consistent with Fig.\ref{dratio}. The frictionless regime must break down at an inertial number $\Ic_c$ where this ratio is of order one, yielding  $\Ic_c \sim \mu_p^2$ in agreement with Fig.\ref{phase}. 

Inside the sliding regime, we have $\epsilon_v \approx \Ic$. To determine the transition to a rolling regime, we note from Eq.~(\ref{energy2}) that $\Dc_{slid}/\Dc_{coll} \sim 1/(\Ic\Lc^{2}) \sim \mu_p^2\chi^2/\Ic$. We observe that the product $\chi \mu_p$ decays with large $\mu_p$ at fixed $\Ic$ (data not shown). Thus, although the dissipation of each sliding contact grows with $\mu_p$, fewer and fewer contacts slide as $\mu_p$ becomes very large, and the latter effect dominates when $\mu_p$ is large enough. This qualitatively explains the observed non-monotonic behavior with $\mu_p$. 
 
\begin{figure*}[t!] 
\includegraphics[width=\textwidth,clip]{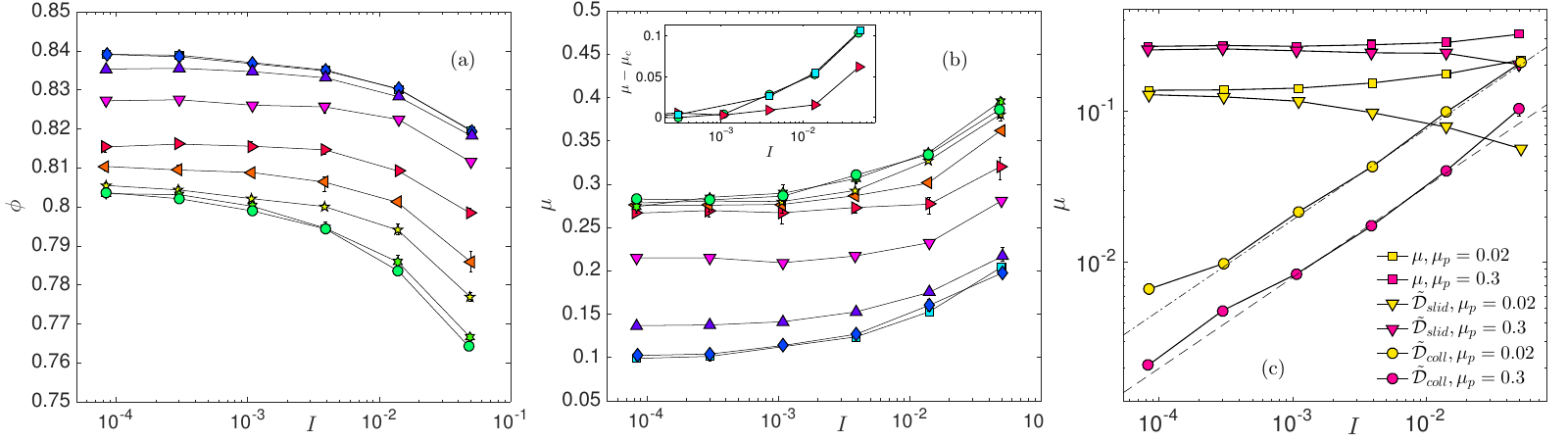}
\caption{(Color online) (a) Volume fraction $\phi$ vs $\Ic$ for various $\mu_p$ (symbols as in Fig. 3). (b) Effective friction $\mu$ vs $\Ic$ (symbols as in Fig. 3). Inset shows non-monotonic behavior of $\mu(\Ic)-\mu_c$, for $\mu_p = 0, 0.3, 10$. (c) Decomposition of $\mu$ into collisional and frictional components. Both lines have slope $0.6$.}\label{phimu}
\end{figure*}

{ \textit{ Constitutive Relations--}}
Experimentally, the most accessible quantities are the constitutive relations $\mu(\Ic)$ and  $\phi(\Ic)$, which we show in Fig.~\ref{phimu}. 
 To discuss universality classes, it would seem appropriate to measure the exponents $\alpha_\mu$ and $\alpha_\phi$ entering Eq.~(\ref{0}). However, these exponents are much harder to measure than those summarized in Table I, because of finite-size effects in the fitting parameters $\mu_c$ and $\phi_c$ \cite{Peyneau08}. Instead, we simply consider the cases $\mu_p=0$, $\mu_p= 0.3$, and $\mu_p=10$ for which our data are respectively in the frictionless, frictional sliding, and rolling regimes. 
 In the inset to Fig.\ref{phimu} we show that $\delta \mu(\Ic) \equiv \mu(\Ic)-\mu_c$ is nearly identical in the frictionless and rolling regimes (the points overlap), and definitely distinct from its behavior in the frictional sliding regime. This observation supports further our claim for three distinct universality classes. 
 
 { We now argue that in Frictional Sliding regime, the exponent $\alpha_\mu$  describing the evolution of the macroscopic friction with inertial number as defined in Eq.(\ref{0}) can be deduced from the exponent $b$ characterizing the velocity fluctuations}. { From Eq.~(\ref{energy}), the partition of dissipation is also a partition of $\mu$. As shown in Fig.\ref{phimu}c, we observe that the contribution from sliding is nearly independent of $\Ic$, while the contribution from collisions is vanishing as $\Ic \to 0$. So long as the variation in sliding dissipation with $\Ic$ is negligible, this implies that $\mu(\Ic)-\mu_c$ is dominated by collisional dissipation, {\it even} in the frictional sliding regime. We find that this is the case for $\Ic\geq 10^{-2.5}$ (data not shown). This facilitates precise measurement of $\mu(\Ic)-\mu_c$ in this range, with which we obtain the measurement $\alpha_\mu = 0.6$ for $\mu_p=0.3$. Moreover, using Eqs.~(\ref{energy},\ref{dcoll2}) we obtain 
 \eq{
\delta\mu(\Ic) \propto \!\frac{\Ic^2 \Lc^2}{\epsilon_v} \! \sim \Ic \Lc^2\sim \Ic^{1-2b},  \qquad & \mbox{Frictional Sliding},
 }
implying $\alpha_\mu=1-2b$. Our prediction $\alpha_\mu\approx 0.6$ is in reasonable agreement with the previous measurement of \cite{Peyneau09} where $\alpha_\mu\approx 0.8$, considering the restricted range of inertial number where we expect this power-law behavior to hold.}

 \begin{figure}[b!] 
\includegraphics[width=0.8\columnwidth,clip]{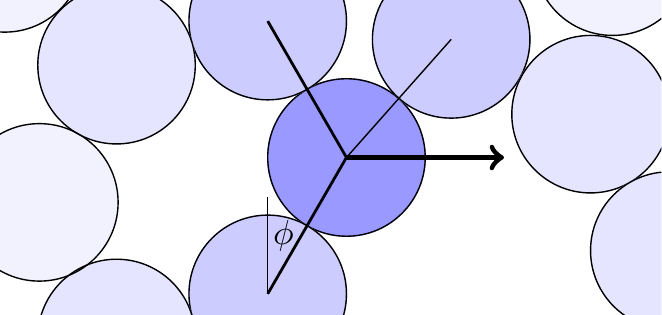}
\caption{(Color online) Illustration of geometrical nonlinearity. If the central grain has an unbalanced force as indicated by the arrow, then the ensuing flow will tend to align the contact normals of the dominant contact forces (thick lines), i.e. the angle $\phi$ will increase. Geometrically, $d\phi/dt \propto V$, the velocity of the particle. }\label{fig_sketch}
\end{figure}


%

{ \textit{Scaling argument for the characteristic strain scale--} }
{ The relation $\epsilon_v \approx \Ic$ can be rationalized by a generalization of the argument in  \cite{DeGiuli15a}. We use the geometrical fact that in dense flows, when a grain has an unbalanced net force, $\Fv$, the ensuing motion will tend to make the remaining contacts of the grain align along $\Fv$ (see Fig.\ref{fig_sketch}). Since forces are repulsive, this further increases the unbalanced force and further accelerates the grain. The increase in net force is proportional to the typical contact force, $p D^{d-1}$, as well as to the rotation of the contacts, of magnitude $\sim \Lc d\epsilon$, thus 
\eq{
\frac{dF}{d\epsilon} \sim p D^{d-1}  \Lc,
}
where $\Lc$ is the dimensionless magnitude of the velocity fluctuation. This equation can also be derived formally, as shown in Appendix 5. In inertial flow, unbalanced forces are proportional to accelerations, $F = p \Ic^2 d\Lc/d\epsilon$, which leads to
\eq{ \label{epsv2}
\frac{d^2 \Lc}{d\epsilon^2} \propto \frac{\Lc}{\Ic^2}.
}
Eq.\ref{epsv2} indicates that there is a characteristic strain scale $\epsilon_v \approx \Ic$ in which a velocity fluctuation grows by an amount proportional to its initial magnitude. In steady flow, such growth must be destroyed by collisions on the same strain scale, since the latter reorganize the direction of particle motion. Hence this is indeed the scale of decorrelation of particle velocities. (At very large $\mu_p$, the direction of a contact force can be unrelated to the contact direction, and corrections to our argument are plausible.)
}

\textit{Discussion--} In this work we have shown that dense inertial granular flows can be classified into three regimes, in a phase diagram spanned by the friction coefficient $\mu_p$ and the distance to jamming, characterized by the inertial number $\Ic$. By considering the microscopic cause of dissipation, we have shown that its nature must change as the friction coefficient $\mu_p$ increases from zero. One eventually leaves the frictionless regime to enter in the frictional sliding regime, where both the kinetics and constitutive relations differ. As $\mu_p$ increases further, fewer contacts slip, and one enters the rolling regime where collisions once again dominate dissipation, and where exponents are consistent with that of the frictionless regime.

Experimentally, these results could be tested by measuring the correlation function $C(\epsilon) = \langle V^y_i(0) V^y_i(\epsilon) \rangle$, which captures both the lever amplitude $\Lc$ (at $\epsilon=0$) and the strain scale $\epsilon_v$. This will require a sufficient resolution in the strain $\epsilon$ that can be probed. Varying the friction coefficient in these studies would also be valuable.


On the theoretical level, a { complete} theory of the frictional sliding regime, the most important in practice, is still lacking. { Here we have proposed a scaling description relating the singularities in the constitutive law  $\mu(\Ic)$  to those in the kinetic observables $\epsilon_v(\Ic)$, $\Lc(\Ic)$ and $\chi(\Ic)$, which can all be expressed in terms of a single unknown exponent $b$.  A key challenge for the future is to predict the value of $b$. Moreover, our arguments are mean-field in nature, as they assume that dissipation occurs rather homogeneously in space, and that velocity fluctuations are described by a single scale $\Lc$. Although there is  evidence that such mean-field arguments are exact for frictionless particles \cite{DeGiuli15a}, they may be only approximate in the frictional case where intermittent strain localization is sometimes reported \cite{Cruz05,Henkes16}. Concerning the rolling regime, why it} has the same scaling exponents as the frictionless regime also needs to be clarified further, beyond their similarity in dissipation mechanism established here. 

Finally, this work could be extended in several directions. It would be very interesting to measure the kinetic quantities presented here in the intermittent quasi-static regime of very slow flows $\Ic\lesssim 10^{-4}$ \cite{Cruz05,Kruyt07,Gaume11,Henkes16}. Similar extensions could be done with respect to particle shape, where local ordering is important \cite{Azema10,Azema12}, and particle softness, where the flow curve can become sigmoidal, leading to hysteresis \cite{Otsuki11,Ciamarra11,Grob14}. Last, over-damped suspensions present the same problem as inertial flows: various numerical studies have focused on frictionless particles \cite{Olsson07,Peyneau09,Olsson11,Heussinger09,Lerner12a,Olsson16}, which appear consistent with the theory developed in \cite{DeGiuli15a}. Together with Eq.~\ref{dslid2}, the theory predicts that frictional sliding should dominate over viscous dissipation when $\eta_0\edot/P \ll \mu_p^2$, where $\eta_0$ is the viscosity of the solvent. It is currently unclear whether this transition qualitatively affects physical properties, as experiments \cite{Boyer11,Dagois15} and numerics \cite{Trulsson12} with friction are reasonably compatible with the frictionless theory. Numerically building a phase diagram analogous to Fig.~\ref{phase}, comparing the amplitude of sliding dissipation to other sources, would resolve this issue.

\begin{acknowledgments}
We acknowledge discussions with M. Cates, G. D\"uring, Y. Forterre, E. Lerner, J. Lin, B. Metzger, M. M\"uller, O. Pouliquen, A. Rosso, and L. Yan. This work was supported primarily by the Materials Research Science and Engineering Center (MRSEC) Program of the National Science Foundation under Award No. DMR-1420073. M.W. thanks the Swiss National Science Foundation for support under Grant No. 200021-165509 and the Simons Collaborative Grant ``Cracking the glass problem".
\end{acknowledgments}
\bibliography{../../bib/Wyartbibnew}

%

\section{Appendix}

In these Appendices, we provide additional details of our results. Appendix 1 describes details of the numerical simulations, while Appendix 2 shows that our main results are independent of the grain stiffness, restitution coefficient, system size, and numerical integration time-step. Appendix 3 shows atypical behavior of sliding velocity and sliding force in the rolling regime. Appendix 4 shows the velocity autocorrelation function for several values of $\mu_p$ and $\epsilon_v$ for all values of $\mu_p$ considered. Appendix 5 computes the effect of geometrical nonlinearity during flow.

\begin{figure*}[ht!] 
\includegraphics[width=\textwidth,clip]{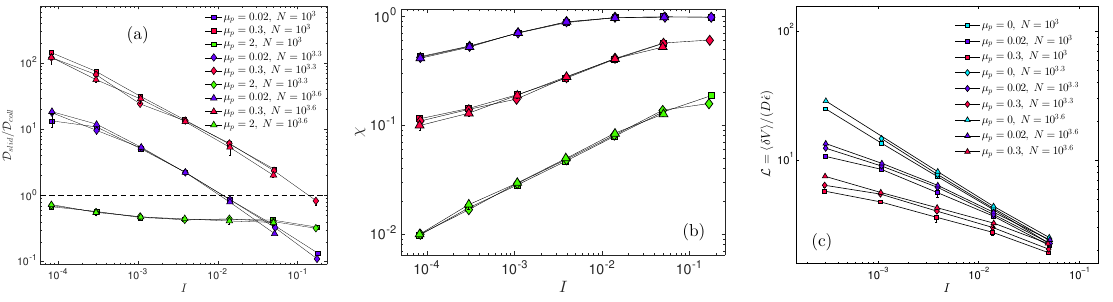}
\caption{(Color online) $N$-dependence of observables. (a) Dissipation ratio vs $\Ic$. (b) $\chi$ vs $\Ic$. (c) $\Lc$ vs $\Ic$. }\label{Ndep}
\end{figure*}
\begin{figure*}[ht!] 
\includegraphics[width=\textwidth,clip]{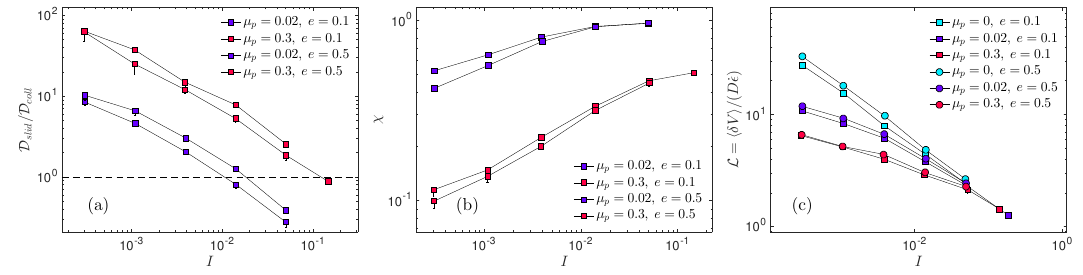}
\caption{(Color online) $e$-dependence of observables. (a) Dissipation ratio vs $\Ic$. (b) $\chi$ vs $\Ic$. (c) $\Lc$ vs $\Ic$. }\label{edep}
\end{figure*}
\begin{figure*}[ht!] 
\includegraphics[width=\textwidth,clip]{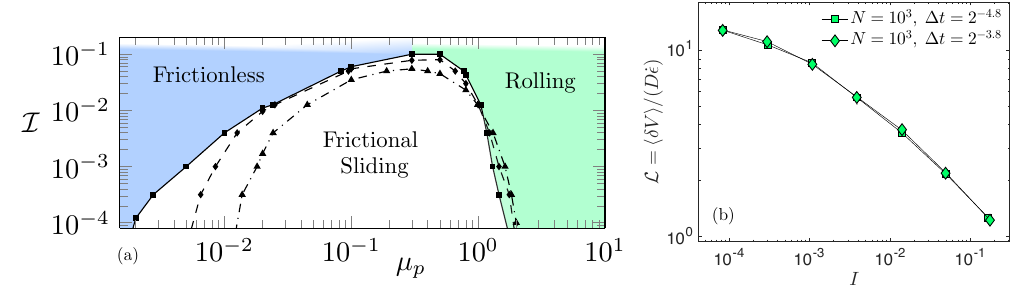}
\caption{(Color online) (a) $\Delta$-dependence of phase diagram for $\Delta =10^{-3.8}$ (squares, solid), $10^{-2.8}$ (diamonds, dashed), and $10^{-1.8}$ (triangles, dash-dotted). (b) Absence of dependence of $\Lc$ on time step $\Delta t$ for $\mu_p=0.02$. }\label{deltadep}
\end{figure*}

\subsection*{\label{si1}Appendix 1. Numerical Simulations}
Simulations are performed with a standard Discrete Element Method code \cite{Cundall79}, which integrates Newton's equations of motion for each grain with Verlet time-stepping. We focus on two dimensions, as empirically exponents do not appear to depend on dimension; see  \cite{DeGiuli15a} for a review of the literature on this point. Collisions are computed by modeling grains as viscoelastic disks: when grains overlap at a contact $\alpha$, they experience elastic and viscous forces $\fv_\alpha^{e}$, and $\fv_\alpha^{v}$. The coefficient of the viscous force is chosen to obtain a restitution coefficient $e=0.1$ in binary collisions; away from the singular limit $e \to 1$ that we do not consider, varying this coefficient does not strongly affect our results, as shown in Section 2. These forces can be decomposed into their contributions normal to the contact, $f_\alpha^{N,e}$ and $f_\alpha^{N,v}$, and tangential to the contact, $\fv_\alpha{}^{\!T,e}$ and $\fv_\alpha{}^{\!T,v}$. 
 The tangential force is imposed to stay inside the Coulomb cone, $|\fv_\alpha{}^{\!T}| \leq \mu_p f_\alpha^N$. Contacts that saturate the Coulomb constraint are said to be {\it sliding}. 
 
The grains are polydisperse with equal numbers of diameter $(0.82,0.94,1.06,1.18)$, the same mixture used in \cite{Grob14}. Previous work established that the polydispersity does not affect $\mu_c$ in simple shear flow, even over a huge range of polydispersity \cite{Voivret09}. Results in the main text are reported for a value of grain stiffness such that the grain relative deformation is set to $\Delta=10^{-3.8}$, appropriate for some materials \cite{Andreotti13}. This is well within the range $\Delta < 10^{-3}$ in which rheological results are independent of $\Delta$, as previously established in simulations of inertial flow of frictionless and frictional particles \cite{Cruz05,Peyneau08,Peyneau09,Trulsson12,Bouzid13}. This result is verified by the explicit dependence of our phase diagram on $\Delta$, reported in Section 2 for $\Delta \in \{ 10^{-3.8}, 10^{-2.8}, 10^{-1.8} \}$. We studied three system sizes $N \in \{1000, 1800, 3700 \}$. Results are reported for the largest $N$; the absence of finite-size effects is established in Section 2. The shortest time-scale of the dynamics is the microscopic elastic time-scale $\sqrt{m/k}$; we chose our numerical time-step $\Delta t$ such that it never exceeds $0.06 \sqrt{m/k}$. This ensures that binary collisions are resolved with $>15$ steps, and the much slower multi-body collisions typical of dense flow will be resolved in even greater detail. Independence of our results with respect to $\Delta t$ is shown in Appendix 2.
 
The square domain of size $L_x \times L_y$ is periodic in the $x$-direction and has upper and lower walls. The walls are created from the same polydisperse mixture as the bulk, staggered to create roughness. The walls obey an equation of motion
\eq{ \label{wall}
M\frac{d^2\rv}{dt^2} + \eta \frac{d\rv}{dt} = \Fv_{bulk} + \Fv_{ext},
} 
where $M$ is mass, $\eta$ is a damping coefficient, $\Fv_{bulk}$ is the force from the bulk of the packing, and $\Fv_{ext}$ is an external applied force. The bulk-wall interactions are via contact forces, exactly as in the bulk. The external force in the $y$-direction is constant, such that $F^{\pm y}_{ext} = \mp P L_x$ on the top (+) and bottom (-) walls. In the $x$-direction, the external force is chosen to impose a constant velocity $\pm V_w$, and hence a constant global shear rate $\edot = 2V_w/L_y$, up to fluctuations in $L_y$. 

We seek to make the flow as homogeneous as possible. Following \cite{Cruz05}, we set $\eta=\sqrt{m k}$, where $k$ is the spring constant for particle-particle elastic interactions, and $m$ the mean particle mass. We tested the dependence of the results on $M$. When $M/m \sim 1$, the wall equation Eq.\eqref{wall} is dominated by the viscous term, and can exhibit long transients. We therefore set $M/m=50$, so that the wall density and particle density are the same order; this minimized transients.

With this choice of wall parameters, we find that steady states are achieved where the relative pressure fluctuations range from 1\% at $\Ic \sim 10^{-5}$ to 20\% at $\Ic \sim 0.1$; thus the mean particle overlap $\Delta \propto P/k$ is fixed to within this precision. 

To prepare homogeneous steady states, initially isotropic packings are created from a gas at volume fraction $\phi_0=0.5$, and then sheared for a pre-strain $\epsilon_0$. As discussed below, an analysis of Eq.\ref{wall}, leads to our choosing $\epsilon_0 = \max(0.2,\Delta^{-1/2} \Ic)$, which we checked ensures that a steady state is reached. After this initial strain, without collecting data, we strain the systems for $\epsilon = 0.3$, collecting data every $\delta \epsilon = 5 \times 10^{-4}$. 

In all cases, we discard runs that are not sufficiently homogeneous. As a first criterion, we exclude simulation runs where the mean velocity profile has a shear band. As a second criterion, we find for certain parameter values that resonant elastic waves bounce back and forth between the walls at very high frequency, as discussed in \cite{Trulsson13}. Resolution of these waves requires a much smaller time step than is needed otherwise, so we do not include these runs. Details of these criteria follow.

To determine an appropriate pre-strain scale $\epsilon_0$, consider the $y$-direction bulk-wall force on on the top wall, $F^{+y}_{bulk}$. This is a spring-like force, since it results from the elastic interactions between the particles adjacent to the wall, and the wall itself, but with a nontrivial spring constant. It can be estimated from the law $\phi_c-\phi \propto \Ic^{\alpha_\phi}$. Indeed, linearizing this law around a mean volume fraction $\bar{\phi}$ and mean pressure $\bar{P}$, we find $P-\bar{P} \propto -\edot^2 (\phi_c-\bar{\phi})^{-1/\alpha_\phi-1} (y_w-L_y/2)/L_y$, where $L_y$ is the mean thickness of the domain and we used $(\phi-\overline{\phi})/\bar{\phi} = -(y_w-L_y/2)/(L_y/2)$. Hence the bulk-wall force is approximately $F^{+y}_{bulk} \sim P L_x \sim -k_w (y_w-L_y/2)$ with $k_w \sim \edot^2 (\phi_c-\bar{\phi})^{-1/\alpha_\phi-1 } \sim \bar{P} (\phi_c-\bar{\phi})^{-1}$. The strain scale associated with the damping term in Eq.\eqref{wall} is then $\epsilon_0 \sim \edot \eta/k_w \sim \Delta^{-1/2} \Ic (\phi_c-\bar{\phi})$, where $\Delta=p/k$. We conservatively take $\epsilon_0 = \max(0.2,\Delta^{-1/2} \Ic)$. 


Our two criteria for ensuring homogeneity of the flows are that there is no static shear band, and that the walls are not in resonant motion. To test for a shear band, we compute the deviation of the mean velocity profile from a linear one, $\delta v(y) = v(y)-\edot y$, and compute its normalized standard deviation, $\langle (\delta v(y)-\langle \delta v \rangle)^2 \rangle_y/(L_y \edot)^2$. For a perfect shear band, this is $1/\sqrt{12} = 0.29$; we discard runs where it exceeds 0.2.

For certain parameter values, resonant elastic waves bounce back and forth between the walls at very high frequency, as discussed in \cite{Trulsson13}. Resolution of these waves requires a much smaller time step than is needed otherwise, and in our code they display an unphysical alternation of the velocity of the wall from positive to negative values at each strain increment where we save data. Therefore we compute a normalized numerical derivative of the vertical wall velocity, $O= (V_w(\epsilon+\Delta \epsilon)-V_w(\epsilon))/(\Delta \epsilon \langle \delta V \rangle)$, where $V_w$ is the wall velocity (for brevity, here we include only one wall), and $\langle \delta V \rangle$ is the velocity scale of grains in the bulk, computed from their fluctuations. We find that for well-behaved runs, $O \Ic \sim 1$, while for numerically unstable ones, $O \Ic > 10000$. Therefore we exclude runs with $O\Ic > 2000$. We checked that the few runs so excluded agree in their location in parameter space with the theory of \cite{Trulsson13}.

\subsection*{\label{si2}Appendix 2. Dependence of results on $\Delta, e, N,$ and $\Delta t$}

In the main text, we reported results for grain relative deformation $\Delta \approx 10^{-3.8}$, number of particles $N \approx 3700$, restitution coefficient $e=0.1$, and time step $\Delta t \approx 0.06 \sqrt{m/k}$. Here we discuss how our results depend on these choices. We show representative plots for $N \approx 1000, 1800, 3700$ at several values of $\mu_p$ and several quantities in Fig.\ref{Ndep}. We see that $\chi$ and $\Dc_{slid}/\Dc_{coll}$ are independent of $N$ at the two largest values studied. Since velocity fluctuations are suppressed at the wall, $\Lc$ displays an expected mild, systematic dependence. This behavior is representative for all values of $\mu_p$. In all cases the minor dependence in $\Lc$ does not affect the scaling behavior of observables, and in particular the phase diagram is not affected. Similarly, representative plots for $e=0.5$ in small systems $N \approx 1000$ show that $\Dc_{slid}/\Dc_{coll}, \chi$, and $\Lc$ display only a weak dependence on the restitution coefficient. 

Although the grain relative deformation $\Delta \approx 10^{-3.8}$ is well within the rigid limit established in previous work \cite{Cruz05,Peyneau08,Peyneau09,Trulsson12,Bouzid13}, we checked how our phase diagram depends on $\Delta \in \{ 10^{-3.8}, 10^{-2.8}, 10^{-1.8} \}$, as shown in Fig.\ref{deltadep}a. For the two smallest values of $\Delta$, the frictionless--frictional-sliding transition is independent of this value above the quasistatic regime, and the frictional-sliding--rolling transition displays only a very small dependence. 

Finally, in Fig.\ref{deltadep}b we show independence of our results on the time-step $\Delta t$, which was halved for a set of simulations with $\mu_p=0.02$, which includes both frictionless and frictional sliding regimes.

\subsection*{Appendix 3. Sliding dissipation in rolling regime}

\begin{figure*}[htb!] 
\includegraphics[width=0.70\textwidth,clip]{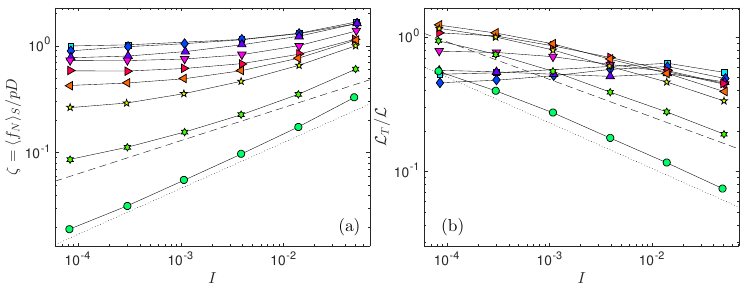}
\caption{(Color online) (a) $\zeta$ vs $\Ic$. Dotted and dashed lines have slopes 0.43 and 0.31, respectively. (b) $\Lc_T/\Lc$ vs $\Ic$. Dotted and dashed lines have slopes -0.34 and -0.28, respectively. }\label{coord}
\end{figure*}

In the rolling regime, only a small subset of contacts are sliding, as shown in Fig.5b of the main text. This raises the possibility that the forces and velocities at these contacts may be atypical of the system as a whole. We investigate this with the quantities $\zeta \equiv \langle f^N \rangle_{S}/(p D^{d-1})$, and $\Lc_T = \langle |\Uv^T| \rangle_{S} /(D\edot)$, where $\langle \cdot \rangle_S$ denotes an average over sliding contacts. As shown in Fig.\ref{coord}, atypical behavior of sliding contacts is indeed shown for large $\mu_p$, and in fact we find in this regime that both $\zeta$ and $\Lc_T/\Lc$ show power-law behavior. In particular, for $\mu_p = 2$ we find $\zeta \sim \Ic^{0.31}$ and $\Lc_T/\Lc \sim \Ic^{-0.28}$, while for $\mu_p = 10$ we find $\zeta \sim \Ic^{0.43}$ and $\Lc_T/\Lc \sim \Ic^{-0.34}$. The quantities $\zeta$ and $\Lc_T/\Lc$ would be needed for an accurate scaling estimate of sliding dissipation in the rolling regime. 


\subsection*{Appendix 4. Velocity autocorrelation function}
\begin{figure*}[tbh!] 
\includegraphics[width=\textwidth,clip]{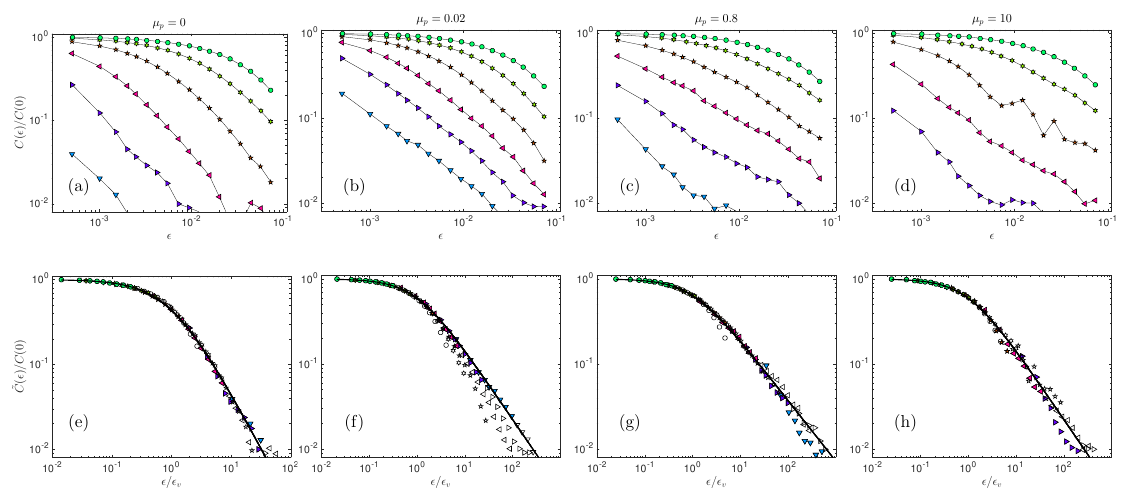}
\caption{(Color online) Autocorrelation of particle velocities, $C(\epsilon) = \langle V^y_i(0) V^y_i(\epsilon) \rangle$ for indicated $\mu_p$ and $\Ic$ from $10^{-4.1}$ (down triangles) to $10^{-1.3}$ (circles) (symbols as in Fig 4a of the main text). (a-d) $C(\epsilon)/C(0)$ vs $\epsilon$. (e-h) $C(\epsilon)/C(0)$ vs $\epsilon/\epsilon_v$. In (e-h), unfilled symbols correspond to strains larger than $0.01$, not used for fitting, and the solid line shows the fitted form. }\label{corrapp}
\end{figure*}
\begin{figure}[tbh!] 
\includegraphics[width=\columnwidth,clip]{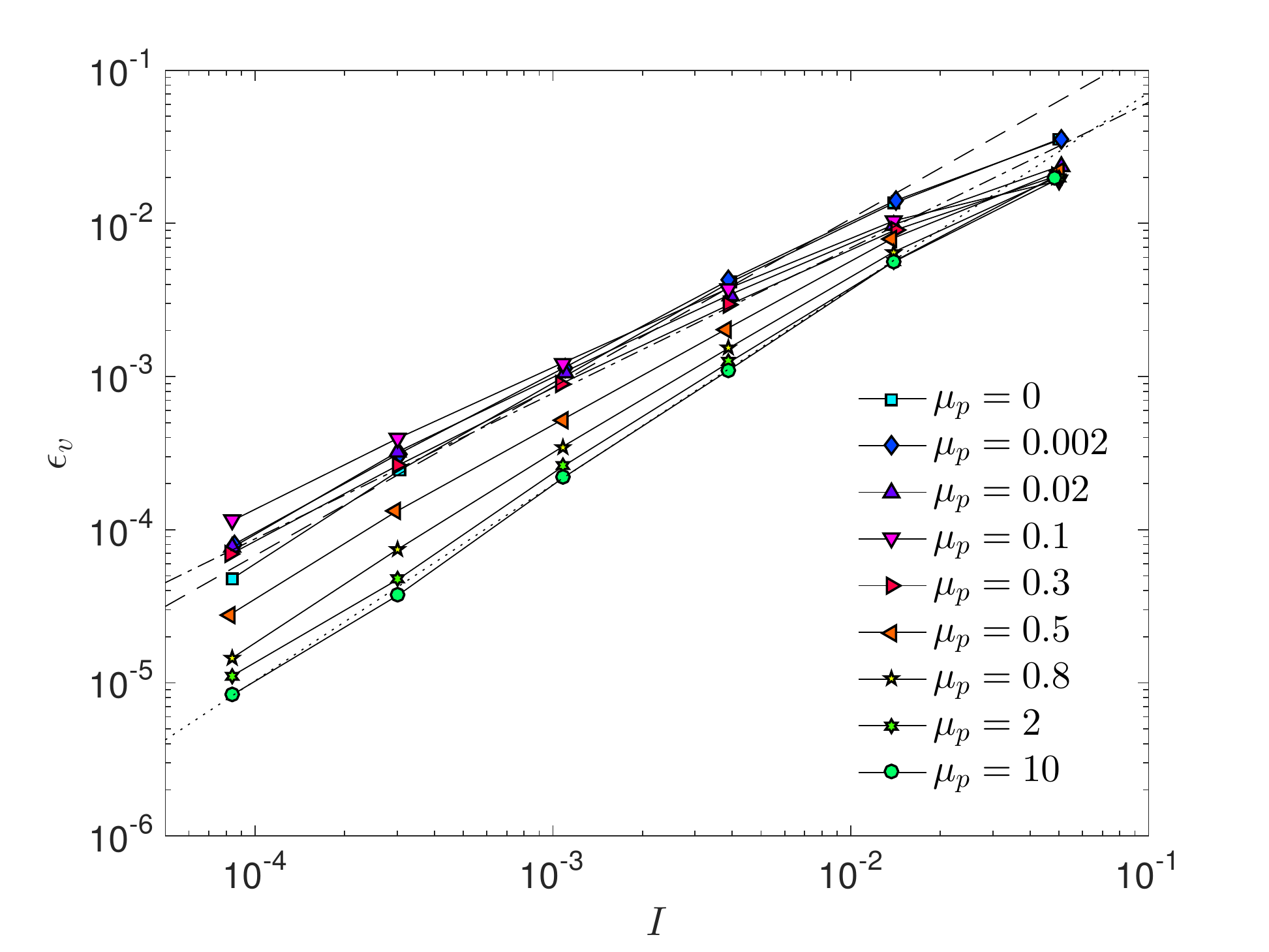}
\caption{(Color online) Strain scale $\epsilon_v$ from collapse of $C(\epsilon)$. Lines as in Fig.5a of the main text.}\label{epsv_app}
\end{figure}
In the main text we introduced the autocorrelation function 
\eq{
C(\epsilon) = \langle V^y_i(0) V^y_i(\epsilon) \rangle,
}
in terms of the vertical component of velocity at particle $i$, $V^y_i$, averaged over all particles and all initial time steps for a given strain increment $\epsilon$. In Fig.\ref{corrapp}, we show $C(\epsilon)/C(0)$ for several values of $\mu_p$. The values of $\eta$ are listed in Table \ref{tab2}, and the resulting values of $\epsilon_v(\Ic)$ are shown in Fig.\ref{epsv_app}.

\begin{table}[h!]

\begin{tabular}{ c || c |c |c |c |c |c |c |c |c }
$\mu_p$ & 0 & 0.002 & 0.02 & 0.1 & 0.3 & 0.5 & 0.8 & 2 & 10 \\ 
\hline
$\eta$ & 1.2 & 1.2 & 0.75 & 0.5 & 0.5 & 0.55 & 0.65 & 0.75 & 0.75 \\
\end{tabular}

\caption{ \label{tab2} Exponent $\eta$ in fit of $C(\epsilon)/C(0)$. }
\end{table}

 \begin{widetext}

\subsection*{Appendix 5. Geometrical nonlinearity in flow}
We aim to compute how forces evolve along flow of hard particles. In particular, we consider flow along a floppy mode, where the relative velocity at a contact, $\uv_{ij}$, is zero at rolling contacts, and only transverse at the sliding contacts. The definition of $\uv_{ij}$, in particular the motion of $j$ relative to $i$ at their mutual contact, is 
\eq{
\uv_{ij} = \Vv_j - \Vv_i + (-1)^d \Delta \omv_{ij} \times \nv_{ij}, 
}
where $\Delta \omv_{ij} = R_j\omv_j + R_i \omv_i$. This holds both in $d=2$ and $d=3$, where the cross product in 2D is defined as $\vec{v} \times \vec{u} = v_i \epsilon_{ij} u_j$ in terms of the Levi-civita symbol $\epsilon_{12}=-\epsilon_{21}=1, \epsilon_{11}=\epsilon_{22}=0$. In 2D $\omv$ becomes a scalar, and $\omv \times \nv = \omega  \times \nv$ is the vector $\omega \epsilon_{kl} \nv_{l}$. In this way we can handle both cases simultaneously. 

By multiplying $\uv_{ij}$ along an arbitrary set of virtual forces and torques, we obtain the theorem of complementary virtual work \cite{Kruyt03}:
\eq{ \label{vwork2a}
- \sum_{ij \in C_S} \uv_{ij}^T \cdot \fv^T_{ij} & = -\sum_i \left[ \Vv_i \cdot \Fv_i + \omv_i \cdot \tauv_i \right] + \sum_{ij \in \p \Omega} \Uv_{ij}^{ext} \cdot \fv_{ij},
}
where $\Fv_i = -\sum_{\ij} \fv_{ij}$ and $\tauv_i = -\sum_{\ij} R_i \nv_{ij} \times \fv_{ij}$ are the net virtual contact force and contact torque on particle $i$, and $\Uv_{ij}^{ext} = -\Vv_i + (-1)^d R_i \omv_i \times \nv_{ij}$ on the boundary. 
 Eq. \eqref{vwork2a} applies only in between collisions. It is important to stress that Eq. \eqref{vwork2a} holds {\it for any virtual force field} $\{ \fv_\alpha \}$. To compute the nonlinearity in flow, we will use the virtual work theorem with $\{ d\fv_\alpha/d\epsilon \}$ as the virtual `forces'. This allows us to identically remove the leading order terms in flow and consider only those that evolve with strain. The basic equation is then
\eq{ \label{nonlin1}
\sum_{\alpha \in \p \Omega} \Uv_{\alpha}^{ext} \cdot \frac{d\fv_{\alpha}}{d\epsilon} & = -\sum_{ij \in C_S} \uv_\alpha \cdot \frac{d\fv_\alpha}{d\epsilon} + \sum_i \left[ \frac{d\Fv_i}{d\epsilon} \cdot \Vv_i +  \frac{d\tauv_i}{d\epsilon} \cdot \omv_i \right] + \sum_i  \omv_i R_i \cdot \sum_{\alpha \sim i}   \frac{d\nv_{\alpha}}{d\epsilon} \times \fv_{\alpha},
}
where contacts are denoted as $\alpha$. Here the last term is needed to precisely cancel the $d\nv_\alpha/d\epsilon$ terms that appear in $d\tau_i/d\epsilon$ when expanded. Under constant stress boundary conditions, we can fix all the boundary forces, so that the LHS vanishes. The terms on the RHS can be simplified using the definition of floppy modes, and the fact that at sliding contacts we have $|\fv^T_\alpha|=\mu_p f^N_\alpha$. After a long computation, in $d=2$ we can rewrite Eq. \eqref{nonlin1} {\it exactly} as
\eq{
0 & = \sum_{\alpha \in C_S} \left[ \mu_p u_\alpha \frac{df^N_\alpha}{d\epsilon} - \frac{f^N_{\alpha} }{\edot \delta r}  (1-\mu_p) u_{\alpha}^2 \right] - \sum_{\alpha} \frac{f^N_{\alpha} }{\edot \delta r}   \big| \Delta \omv_{\alpha} \big|^2 + \sum_i \left[ \frac{d\Fv_i}{d\epsilon} \cdot \Vv_i +  \frac{d\tauv_i}{d\epsilon} \cdot \omv_i \right],
}
where $u_\alpha$ is the magnitude of sliding velocity at contact $\alpha$. In $d=3$ there are several additional terms that are not expected to be important, for example involving the slight difference between sliding directions and the directions of tangential forces.

When flow is along floppy modes, velocities have a characteristic scale $\delta V$; we will assume that the angular and linear velocities have the same scale, $\delta \omega \sim \delta V/D$. Then since $\chi \lesssim 1$, in terms of scaling we have
\eq{
0 \approx \chi N_C \mu_p \delta V \frac{dp}{d\epsilon} - N_C \frac{p}{\edot} \delta V^2 + N \frac{dF}{d\epsilon} \delta V,
}
where we used that $d\Fv_i/d\epsilon \cdot \Vv_i > 0$. Under constant stress BCs, the $dp/d\epsilon$ term is negligible (more precisely, it vanishes up to correlations between $u_\alpha$ and $df^N_\alpha/d\epsilon$). Then we find
\eq{ \label{dFde}
\frac{dF}{d\epsilon} \approx \frac{z}{2} \frac{p}{\edot} \delta V,
}
as stated in the main text. Eq.\eqref{dFde} indicates how quickly configurations flow out of equilibrium along floppy modes, and applies for both viscous and inertial dynamics. The magnitude of unbalanced forces, $F$, can itself be written in terms of geometrical quantities, but this depends on the dynamics.
 \end{widetext}


\end{document}